\pdfoutput=1
\documentclass[a4paper,
               ]{jacow}
\usepackage{pdfpages,multirow,ragged2e} %
\makeatletter%
	\ifboolexpr{bool{xetex}}
	 {\renewcommand{\Gin@extensions}{.pdf,%
	                    .png,.jpg,.bmp,.pict,.tif,.psd,.mac,.sga,.tga,.gif,%
	                    .eps,.ps,%
	                    }}{}
\makeatother

\ifboolexpr{bool{xetex} or bool{luatex}} 
 {}                                      
 {\usepackage[utf8]{inputenc}}           

\usepackage[USenglish]{babel}

\ifboolexpr{bool{jacowbiblatex}}%
 {%
  \addbibresource{jacow-test.bib}
  \addbibresource{biblatex-examples.bib}
 }{}
\usepackage{amsmath}
\usepackage{multicol}

\usepackage{placeins}
\usepackage{hyperref}
\hypersetup{pdfauthor={Leonard Thiele et al.},pdftitle={Beam-cavity interactions in the rapid cycling synchrotron chain of the future muon collider}}
\begin{document}

\pagenumbering{arabic}

\title{Beam-cavity interactions in the rapid cycling synchrotron chain of the future muon collider}

\author{L. S. Thiele \thanks{leonard.thiele@cern.ch}, U. van Rienen \textsuperscript{1}, University of Rostock, Rostock, Germany \\
		F. Batsch, R. Calaga, H. Damerau, A. Grudiev, I. Karpov, CERN, Geneva, Switzerland \\
		\textsuperscript{1}also at Department of Life Light and Matter, University of Rostock, Rostock, Germany}
\maketitle

\begin{abstract}
   The International Muon Collider Collaboration (IMCC) is engaged in a design study for a future facility intended to collide muons. Subsequent to the initial linear acceleration, the counter-rotating muons and anti-muons are accelerated in a chain of rapid cycling synchrotrons (RCS) up to the multi-TeV collision energy. To maximise the number of muons available in the collider, it is essential to exploit the time dilation of the muon lifetime by employing a large accelerating gradient.
   The 1.3 GHz TESLA cavity serves as the baseline for the RCS chain. Considering the high bunch population and the small aperture of the cavity, the resulting beam-induced voltage per bunch passage is considerable, resulting in a substantial perturbation of the cavity voltage for subsequent bunch passages. In this contribution, the effects of beam loading during the acceleration cycle on the muons are calculated with the objective of determining the optimum parameters for minimising the cavity voltage transients. The interaction of the induced voltages, considering the counter-rotating beams, is studied. 
\end{abstract}
\section{Introduction}
At the low-energy side of the muon collider facility, a superconducting linear accelerator provides a proton beam for the production of the muons on a graphite target. 
The transverse emittance of the beam is thereafter reduced in a series of muon cooling channels. 
Subsequently, a chain of medium-energy linear accelerators provides a pre-acceleration to an energy of \SI{63}{GeV}. 
The following high-energy acceleration is achieved with a cascade of four rapid-cycling synchrotrons (RCS), equipped with large-scale superconducting radio-frequency (SRF) systems. 
Two bunches, one consisting of $\mu^+$ and the other of $\mu^-$ particles, will be accelerated in counter-rotation, sharing the same beam pipe and cavities. 
The parameters of the high-energy acceleration chain in the Muon Collider greenfield study are summarized in Table \ref{tab:mucol_accel_params}. \\
\begin{table}
   \centering
    \setlength{\tabcolsep}{3pt}
    \begin{tabular}{lcccc}
    \toprule
    & \textbf{RCS1} & \textbf{RCS2} & \textbf{RCS3} & \textbf{RCS4} \\
    \midrule
          Circumference [m] & 5990 & 5990 & 10700 & 35000 \\ 
          Injection energy [TeV] & 0.06 & 0.31 & 0.75   & 1.50 \\
          Ejection energy [TeV] & 0.31 & 0.75 & 1.50   & 5.00 \\  
          Acceleration time [ms]  & 0.34 & 1.10 & 2.37   & 6.4 \\
          $\Delta E$ per turn [GeV] & 14.8 & 7.9 & 11.4   & 64 \\
          $G_\mathrm{avg}$ [MV/m] & 2.44 & 1.33 & 1.06 & 1.83 \\
          Harmonic number [\num{E3}] & 26 & 26 & 46   & 151 \\
          Bunch intensity [$10^{12}$] & <2.7 & <2.43 & <2.2 & <2.0 \\ 
          Bunch length [$1\sigma$, ps] &  <31 & <30 & <23 & <13 \\
       \bottomrule
   \end{tabular}
    \caption{Acceleration parameters for the chain of four rapid cycling synchrotrons (RCS) in the muon collider greenfield study, assuming a $90\, \%$ survival rate per RCS \cite{parameter_report}.}
   \label{tab:mucol_accel_params}
\end{table}

\begin{figure}
    \centering
    \includegraphics[width=\linewidth]{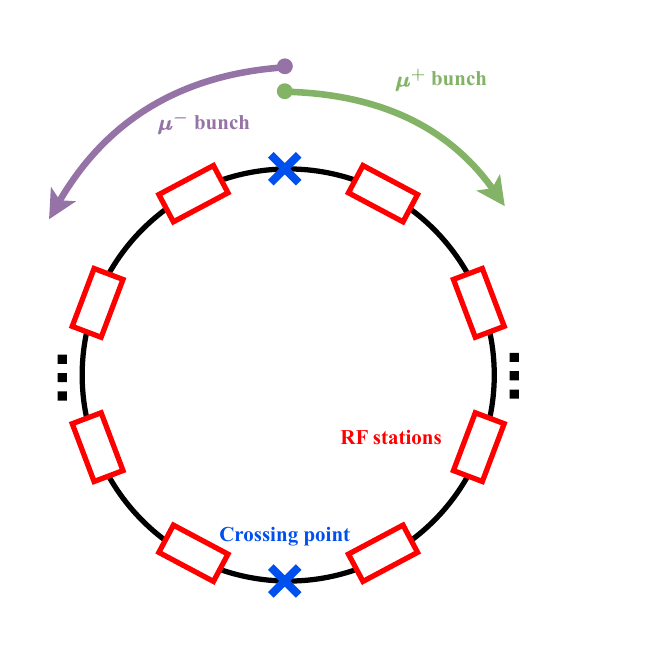}
    \caption{Basic layout of the radio-frequency (RF) system in the rapid cycling synchrotrons. } 
    \label{fig:rf_system_layout}
\end{figure}

One of the main challenges for the design of the RF system is the short muon lifetime of $\tau_\mu=\gamma \, \times\, 2.2\, \mu s$, with the Lorentz factor, $\gamma$. 
In order to quickly increase the energy and benefit from the time dilation through an increased relativistic $\gamma$, a high average gradient is required in the acceleration chain. \\ 
Existing high-gradient RF cavities were evaluated for their suitability for the acceleration of muons. The TESLA cavity type at \SI{1.3}{GHz} \cite{SC_TESLA_cavity} is assumed as a baseline and was demonstrated to be operational at gradients of \SI{30}{MV/m}. 
\begin{table}
    \centering
    \begin{tabular}{lc}
    \toprule
        Parameter & Value \\ \midrule
        Resonance frequency [GHz] & 1.3 \\
        Active length [m]& 1.038 \\
        Cavity voltage [MV] & 31.13 \\
        Geometric shunt impedance [$\Omega$] & 518\\
        RF to wall plug efficiency [\%] & $\sim$48\\
        \bottomrule
    \end{tabular}
    \caption{TESLA cavity parameters \cite{SC_TESLA_cavity} and ILC RF powering system efficiency in the distributed Klystron scheme \cite{ILC_tdr}.}
    \label{tab:TESLA_cavity_values}
\end{table}
This cavity type has been extensively studied and well-optimized. Parameters for the higher-order modes and the fundamental mode are precisely known and validated with measurements. 
The cavity parameters used in this contribution are shown in Table \ref{tab:TESLA_cavity_values}. 
To ensure a stable acceleration, the RF voltage in the cavities in the RCS chain has to be kept constant over tens of turns. 
At the same time, the acceleration process should be as energy-efficient as possible, provided the parameters of the different accelerators. 
Therefore, the requirements for the RF system can be considered as a combination of the typical demands associated with linear accelerators, i.e., high instantaneous power, with those typically needed in a synchrotron, i.e., turn-by-turn voltage stability in the cavity. \\\\
The very high synchrotron tune, $Q_s$ of up to $\approx 0.5$ due to the huge RF voltage per turn requires splitting the RF system into multiple stations around the ring \cite{hb_batsch}. 
In the first RCS, 32 RF stations are foreseen, while the following RCSs require about 24 RF stations to distribute the longitudinal kick and lower $Q_s$ sufficiently. 
If the $Q_s$ is not reduced, a stable, longitudinal emittance preserving acceleration is not possible, thus reducing the collider's available luminosity. 
Consequently, the topology of the RF system in the rings will resemble the layout illustrated in Fig. \ref{fig:rf_system_layout}. 
\subsection{Equivalent circuit modelling of RF systems}
\begin{figure}
    \centering
    \includegraphics[width=0.8\linewidth]{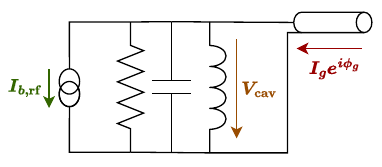}
    \caption{RLC circuit representation of the beam-cavity interaction with the beam current, $I_{b\mathrm{, rf}}$, the cavity voltage, $V_\mathrm{cav}$ and the generator current, $I_g$.}
    \label{fig:rlc_circuit_model}
\end{figure}

{
When designing an accelerating RF system, special attention is paid to the compensation of the beam loading in the fundamental mode. 
While the RF system is designed to deliver energy to the beam, the beam additionally induces a voltage, $V_\mathrm{ind}$ during its passage. 
Ideally, the generator current should fully compensate for both effects to ensure stable acceleration if the beam were to be distributed around the ring. 
The interaction between the beam and the generator current can be modelled by an equivalent RLC circuit model, presented in Fig. \ref{fig:rlc_circuit_model}. 
The values for the resistor $R$, inductor $L$, and capacitor $C$ lumped elements can be calculated from RF and beam parameters and allow the circuit to be analysed in transient and steady-state \cite{cavity-beam-transmitter}.
}

\subsection{RF system parameters in the RCS chain}
With the aforementioned $RLC$ approximation, a set of parameters for the minimum power requirement can be determined by adjusting both the cavity detuning, $\Delta\omega$ and the loaded quality factor, $Q_L$ \cite{cavity-beam-transmitter}. 
The power injected into the cavity will then be fully absorbed by the beam, and the field decay, with no power being reflected back to the generator. 
Any reflected power would have to be directed into a load to protect the generator and, therefore, be lost. 
The optimum loaded quality factor can be determined with \cite{tbeamloading}
\begin{equation}
    Q_{L\mathrm{, opt}} = \frac{V_\mathrm{cav}}{(R/Q)|F_b|I_{b\mathrm{, DC}}\cos{\phi_s}},
\end{equation}
while the optimum cavity detuning is calculated by 
\begin{equation}
    \Delta\omega_\mathrm{opt} = -\frac{(R/Q)|F_b|I_{b\mathrm{, DC}}\sin{\phi_s}\omega_\mathrm{rf}}{2V_\mathrm{cav}}, 
\end{equation}
with the angular RF frequency $\omega_\mathrm{rf}$, the cavity voltage $V_\mathrm{cav}$, the accelerating synchronous phase $\phi_s$, the geometric shunt impedance $R/Q$, the average beam current $I_{b\mathrm{, DC}}$ and the bunch form factor $F_b$. 
Bunches must be kept short, hence a worst-case form factor of $F_b = 2$ is assumed. 
The average beam current is given by
\begin{equation}
    I_{b\mathrm{, DC}} = N_p e f_\mathrm{rev} N_b
\end{equation}
with the number of particles, $N_p$, the elementary charge, $e$ and the number of bunches, $N_b$. 
The synchronous phase, $\phi_s$ is defined in the electron-accelerator convention with \SI{0}{\textdegree} being on-crest. \\
In addition to the power consumption, cavity detuning and the loaded quality factor also affect the beam stability. 
The Robinson high-intensity stability criterion states that the relative beam loading, $Y$ has to satisfy \cite{hbook_accel_physics}
\begin{equation}
    Y=\frac{|F_b|I_{b\mathrm{, DC}}(R/Q)Q_L}{V_\mathrm{cav}} < - \frac{2 \sin{\phi_s}}{\sin{\phi_z}},
\end{equation}
with $\phi_z=\tanh{2Q_L\Delta\omega/\omega_\mathrm{rf}}$ being the detuning angle of the cavity. 
When applying $\Delta\omega_\mathrm{opt}$ and $Q_{L\mathrm{, opt}}$, the system is not Robinson-stable, hence, a larger detuning with $\Delta\omega_m=\Delta\omega_\mathrm{opt}/\sin{\phi_s}$ is chosen \cite{hbook_accel_physics}. 
If a detuning $\Delta\omega\neq\Delta\omega_\mathrm{opt}$ is applied to the cavity, the loaded quality factor, $Q_L$ at which the minimal reflected power occurs is determined according to 
\begin{equation}
\begin{split}
    Q_{L\mathrm{, opt}} = \frac{V_\mathrm{cav}}{(R/Q)}\Biggl\{&\Biggl[\frac{2V_\mathrm{cav}\Delta\omega}{\omega_\mathrm{rf}(R/Q)} + |F_b|I_{b\mathrm{, DC}}\sin{\phi_s}\Biggr]^2 \\
    &  + (|F_b|I_{b\mathrm{, DC}}\cos{\phi_s})^2\Biggr\}^{-1/2}.
\end{split}
\end{equation}
Operating the system under these conditions causes part of the generator power to be reflected by the cavity. 

Table \ref{tab:rf_params} presents the main RF parameters of the muon collider high-energy acceleration. 
While the instantaneous cavity powering requirements are challenging, the duty factors of the machines are comparatively low. 
RCS1 has the lowest duty cycle but the highest RF power per cavity. The latter is caused by a low revolution period and high beam intensity. 
In the subsequent RCSs, the duty cycle rises while the instantaneous power decreases. However, the total average RF power increases due to the large number of cavities employed as well as the increased acceleration time. \\
\begin{table}
    \centering
    \setlength{\tabcolsep}{3pt}
    \begin{tabular}{lcccc}
         \toprule
         & \textbf{RCS1} & \textbf{RCS2} & \textbf{RCS3} & \textbf{RCS4}\\ 
         \midrule
         Beam current [mA] &  43.3&  39&  19.8&   5.49 \\
         Total RF voltage [GV]&  20.9&  11.2&  16.1&   90\\
         Number of cavities &  683&  366&  524&   2933\\
         RF section length [m] &  962&  519&  746&   4125\\ \hline
         Loaded Q-factor [\num{E6}] &  0.696&  0.775&  1.533&   5.522\\
         Cavity detuning [kHz] & -1.32& -1.186& -0.6&  -0.166 \\
         Acceleration time [ms] & 0.34& 1.1& 2.37& 6.37\\
         Cavity filling time [ms] & 0.171& 0.19& 0.375& 1.352\\
         RF duty factor [$\%$] & 0.19& 0.57& 1.22& 3.36\\
         Max. cavity power [kW] & 1128& 1017& 516& 144 \\
         Average power [MW] & 1.919& 2.84& 4.43& 18.92 \\
         Wall plug power [MW] & 2.95& 4.38& 6.811& 29.1\\
         \bottomrule
    \end{tabular}
    \caption{RF system parameters for the greenfield study of the future muon collider \cite{parameter_report}. The calculations assume a \SI{1.3}{GHz} TESLA cavity \cite{SC_TESLA_cavity} as a baseline and a synchronous phase, $\phi_s$ of 45\,\textdegree. Both the average and wall plug power only include the contribution of the RF system.}
    \label{tab:rf_params}
\end{table}
\begin{figure}
    \centering 
    \includegraphics[width=\linewidth]{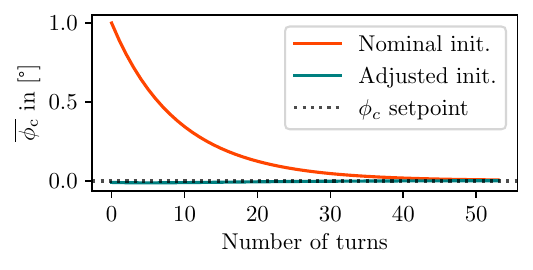}
    \caption{Evolution of the cavity phase, $\phi_c$, averaged over one turn, during the acceleration in RCS2 with the initial parameters set to nominal or adjusted values. RCS2 was chosen, as the acceleration is sufficiently long to make a convergence of the voltages for the nominal initial parameters visible. }
    \label{fig:avg_phase_RCS2}
\end{figure}
\begin{figure*}
    \centering
    \includegraphics[width=\textwidth]{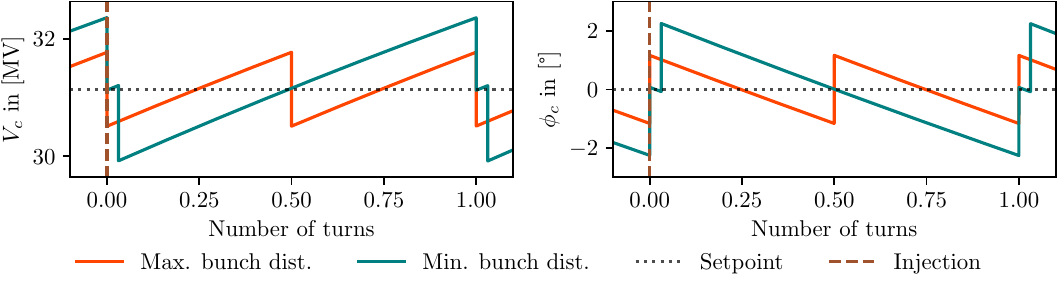}
    \caption{Evolution of the cavity voltage amplitude, $V_c$ (left) and phase, $\phi_c$ (right) during the first turn of RCS1, once considering the maximum and once the minimum bunch distance. The dotted lines represent the respective setpoints. }
    \label{fig:beam_phase_comp}
\end{figure*}
\subsection{Counter-rotating beams}
A complication of the muon collider high-energy chain is the necessity to accelerate two oppositely charged beams in the same beam pipe and through the same cavities simultaneously. 
As both the sign and the direction are reversed, both bunches must be accelerated at the same synchronous phase, $\phi_s$ in the cavities. 
The induced voltages at the fundamental mode will, therefore, constructively interfere. 
The simulation of the voltage evolution can be performed by assuming the same direction with a variable time delay between the bunch passages. 
Note, as the bunches cross at two points in the ring and the RF system is distributed, the arrival times of the bunches will differ from station to station due to the non-equal distances to the crossing point. \\\\
However, a similar approach cannot be applied to simulate the interactions of higher-order modes (HOM). 
Each mode has a different wavelength, which causes the HOM contributions from the counter-rotating beams to interfere constructively or destructively. 
Generally, care should be taken to avoid a HOM resonance frequency at an integer multiple of the revolution frequency, $f_\mathrm{rev}$. 
The HOM would then constructively interfere within each turn, causing an in-phase voltage build-up. 
As a particular case in the muon-collider, additionally, HOMs at frequencies of 
\begin{equation}
\begin{split}
    &f_\mathrm{HOM} \neq n * \Delta_\mathrm{dist}\,/\, N_\mathrm{rf-stations} * f_\mathrm{rev},\\
    \mathrm{with} \, \Delta_\mathrm{dist} &\in \{1, 3, \ldots , N_\mathrm{rf-stations} \, / \, 2 - 1\} \, \mathrm{and} \, n \in \mathbb{N}
\end{split}
\end{equation}
should be minimised in the cavity design. As the bunch separation will differ in every station, in-phase interferences need to be taken into account. 
Especially HOMs with a high loaded quality factor, $Q_L$ and, therefore, long decay time should be avoided. 
The HOM damping scheme and related effects on the beam dynamics are currently under study. 
Should specific resonances prove problematic, it may be necessary to alter the cavity design. 

\section{Transient beam loading}
Since the beam in the muon acceleration chain consists of only two bunches, the no-beam segment between bunch arrivals can be extremely large. 
Therefore, a condition for the steady state of the cavity voltage, achieved in more evenly filled rings, will not be reached. 
An analysis of the evolution of the fundamental mode voltage in the cavity requires the modelling of the transient beam loading. 
In such semi-analytical simulations, it is assumed that the cavity voltage amplitude, $A_c$ and phase, $\phi_c$ vary with:
\begin{equation}
    V_c(t) = A_c(t)e^{i\phi_c(t)}.
\end{equation}
The following two equations describe the change in phase and amplitude of the cavity voltage due to the beam passage and the generator current. They were derived from the $RLC$ circuit model in Fig. \ref{fig:rlc_circuit_model} in \cite{cavity-beam-transmitter}. 
\begin{equation}
\begin{split}
    \frac{d\phi_c(t)}{dt} = \Delta\omega &+ \frac{(R/Q)\omega_\mathrm{rf}}{A_c(t)}\times\Biggl\{I_g\sin[\phi_g- \phi_c(t)]\\
     &+ \frac{A_b(t)\sin[\phi_s - \phi_b(t) + \phi_c(t)]}{2}\Biggr\},
     \label{eq:phi_c}
\end{split}
\end{equation}
\begin{equation}
\begin{split}
    \frac{dA_c(t)}{dt} = -\frac{A_c(t)}{\tau}
    &+ (R/Q)\omega_\mathrm{rf}\times \Biggl\{I_g\cos[\phi_g- \phi_c(t)]\\
    &- \frac{A_b(t)\cos[\phi_s - \phi_b(t) + \phi_c(t)]}{2}\Biggr\}, 
    \label{eq:ampl_c}
\end{split}
\end{equation}
with the cavity voltage filling time constant $\tau=2Q_L\,/\,\omega_\mathrm{rf}$, the generator current, $I_g$, the generator phase, $\phi_g$, and the beam current, $A_b$. 
The amplitude of the beam current is obtained from 
\begin{equation}
    A_{b\mathrm{, peak}} = N_p f_\mathrm{rf}e |F_b|, 
\end{equation}
during RF periods, where the beam is present and vanishes elsewhere. \\\\
As a consequence of the alteration in the amplitude and phase of the cavity voltage, it is necessary for the beam phase, $\phi_b(t)$, to perform an alteration in order to ensure a constant energy gain. 
\begin{equation}
    A_c(t)\cos[\phi_s - \phi_b(t) + \phi_c(t)] = V_\mathrm{cav}\cos{\phi_s},
    \label{eq:beam_phase}
\end{equation}
with $V_\mathrm{cav}$ being the nominal cavity voltage and $\phi_s$ being the nominal accelerating phase. 
Due to the synchrotron motion, the beam phase, $\phi_b$ automatically adjusts to this phase. 
As Eq.~\eqref{eq:phi_c} and Eq.~\eqref{eq:ampl_c} cannot be solved analytically, numerical analysis using the Euler method is performed. \\
The generator current amplitude, $I_g$ and phase, $\phi_g$ are assumed to be constant following 
\begin{equation}
    I_{g}e^{i\phi_g} = \frac{V_\mathrm{cav}}{2(R/Q)}\biggl(\frac{1}{Q_L} - 2i\frac{\Delta\omega}{\omega_\mathrm{rf}}\biggr) + |F_b|I_{b\mathrm{, DC}} e^{-i\phi_s}. 
\end{equation}
This complex current compensates for the beam loading during each turn. 
\subsection{Tuning of initial parameters}
The low number of turns requires a more evolved approach to correctly determine the initial values of the cavity voltage amplitude and the phase at the injection of the beam. 
If the acceleration period would last longer, an equilibrium would be reached, and the cavity phase, $\phi_c$ and cavity voltage amplitude, $A_c$ would oscillate around fixed values during one turn. 
However, to ensure a stable acceleration, this transient behaviour should be avoided. 
The initial setpoint can be estimated by approximating the parameters in the beam gap in order to calculate the voltage change. 
The relative phase between the cavity voltage and the generator can be approximated by $\phi_g$. 
Furthermore, the phase and amplitude of the cavity will, on average, be at their setpoint values, i.e. zero and $V_\mathrm{cav}$, respectively. 
The initial voltage can then be obtained from Eq.~\eqref{eq:ampl_c} by calculating the voltage change in the largest non-beam segment:  
\begin{equation}
\begin{split}
    V_\mathrm{cav, sep} =\,& \biggl(-\frac{V_\mathrm{cav}}{\tau} + (R/Q)\omega_\mathrm{rf}I_g\cos\phi_g\biggr)\times t_\mathrm{sep}\\
    V_\mathrm{cav, init} =\,& V_\mathrm{cav} + V_\mathrm{cav, sep}~ / ~2.
\end{split}
\end{equation}
\noindent
Here, $ t_\mathrm{sep} $ is the largest separation time of the bunches, defined as the time interval without beam interaction, and $ V_\mathrm{cav, sep} $ is the corresponding cavity voltage change during this time. 
Analogously, the initial cavity phase can be calculated from Eq.~\eqref{eq:phi_c} as 
\begin{equation}
    \phi_\mathrm{cav, init} = \biggl(\Delta\omega + \frac{(R/Q)\omega_\mathrm{rf}I_g\sin\phi_g}{V_\mathrm{cav}}\biggr)\times t_\mathrm{sep}~ / ~2.
\end{equation}
Figure \ref{fig:avg_phase_RCS2} shows that the average cavity phase takes a considerable amount of time to reach equilibrium with nominal parameters as initial values. 
When the adjusted initial parameters are chosen, this change is much less pronounced, thus significantly stabilising the cavity condition during acceleration. 
As the phase and amplitude change in the initial turns of the acceleration, the required beam phase and, consequently, the bucket area are reduced, potentially leading to the loss of particles.
\begin{figure}
    \centering
    \includegraphics[width=\linewidth]{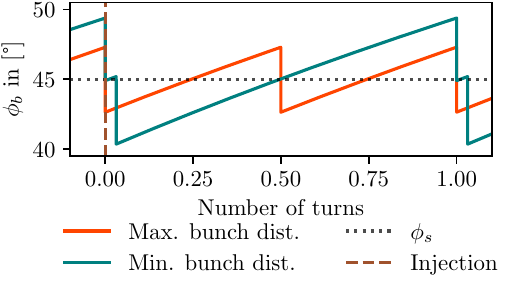}
    \caption{Evolution of the beam phase, $\phi_b$ during the first two turns of RCS1. At a minimum bunch distance, the beam phase has to shift by a larger amount, as both the lowered cavity voltage and phase need to be compensated. 
    The dotted line indicates the synchronous phase, $\phi_s$, the setpoint value of the beam phase, $\phi_b$.}
    \label{fig:beam_phase_dist}
\end{figure}
\subsection{Simulation with different bunch distances}
The time difference between the arrival of the counter-rotating muon bunches will strongly depend on the position of the RF station within the ring. The maximum time separation is given by $t_\mathrm{sep, max} = h\,/\, 2 \times t_\mathrm{rf}$ with the harmonic number, $h$ and the RF period, $t_\mathrm{rf}$. 
This corresponds to an RF station position halfway between the crossing points in Fig.\,\ref{fig:rf_system_layout}. 
Assuming equally spaced RF stations, the minimum separation is dependent on the number of RF stations with $t_\mathrm{sep, min} = h / N_\mathrm{rf-stations}\times t_\mathrm{rf}$, which corresponds to stations directly next to the crossing point. 
Depending on the arrival time difference of the bunches, both the variation of the cavity voltage, $V_c$ and the cavity phase, $\phi_c$ differ significantly, as can be observed in Fig. \ref{fig:beam_phase_comp}. 
RCS1 is chosen for the following comparisons, as the changes are expected to be the most pronounced here. 
The deviation of the cavity voltage from its centre point during the acceleration is of the order of \SI{1}{MV}. 
This leads to the additional requirement to lower the setpoint voltage to avoid a breakdown inside the cavity, which, in turn, increases the total amount of required cavities. \\\\
From Eq.~\eqref{eq:beam_phase} it is evident that the adjustment of the beam phase, $\phi_b$ must be more significant than the cavity phase adjustment, as both the reduction in cavity voltage amplitude and the deviation of the cavity phase need to be compensated. 
The evolution of the beam phase during acceleration is shown in Fig. \ref{fig:beam_phase_dist}. 
The calculated deviation of the beam phase, $\phi_b$ from its design value, $\phi_s$, causes a variation of the bucket area during the acceleration, which, in steady-state can be approximated with \cite{lee_accel_phys} 
\begin{equation}
    A_b(t) \approx \frac{8}{\omega_\mathrm{rf}}\sqrt{\frac{2qA_c(t)\beta_sE_s}{\pi h |\eta|}}\frac{1 - \cos{\phi_b(t)}}{1 + \cos{\phi_b(t)}}.
\end{equation}
Here $\beta_s$ is the relativistic speed of the synchronous particle, $E_s$ the energy of the synchronous particle, $q$ the charge of the particle, and $\eta$ the slippage factor of the accelerator.\\
Particularly in RCS1, where the bucket area is only marginally larger than the bunch emittance \cite{hb_batsch}, detailed beam dynamics simulations are required to verify that no beam loss occurs due to insufficient bucket area. 
Nonetheless, the reduction in the bucket area is symmetric for both bunches and, therefore, does not result in uneven energy gain or asymmetric beam loss. 
\subsection{Simulation in different RCSs}
\begin{figure}
    \centering
    \includegraphics[width=\linewidth]{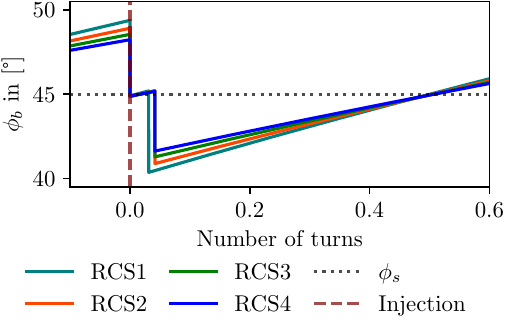}
    \caption{Comparison of the evolution of the beam phase, $\phi_b$ in the different RCSs. The arrival time difference between the bunches is assumed as  $h\,/\,N_\mathrm{rf-stations}\,\times\, t_\mathrm{rf}$. In RCS1, 32 RF stations are present, while RCS2 to RCS4 feature 24 stations.}
    \label{fig:comp_RCS_chain}
\end{figure}

Figure \ref{fig:comp_RCS_chain} compares the magnitude of the adjustments of the beam phase, $\phi_b$ in the different RCSs.
In the higher-energy RCSs, this effect is less pronounced. 
The amplitude depends on the energy gain per cavity, the bunch intensity and the number of RF stations. 
A higher number of RF stations reduces the relative arrival time difference and, therefore, increases the remaining disturbance of the cavity voltage from the counter-rotating bunch. 
The higher bunch intensity in the smaller RCSs results in more energy being extracted from the cavity during the bunch passage, resulting in a larger voltage drop. 
The generator current, $I_g$, the loaded quality factor, $Q_L$ and the cavity detuning, $\Delta\omega$ are adapted to the beam and cavity parameters and ensure that the voltage oscillates around the setpoint. 
Should the magnitude of the beam phase change prove problematic in beam dynamics studies, the energy gain per cavity could be reduced, in order to lower the required beam phase and increase the bucket area. 
Alternatively, the generator current has to be altered during the acceleration to compensate for the asymmetric bunch arrival times. 
\section{Conclusion}
The main challenge of the RF system design in the RCS chain of the future muon collider is to ensure that RF voltage for a sufficiently fast energy gain is available during the whole acceleration process. 
This contribution presents the results of a transient beam loading simulation with a static generator current. 
Due to the large beam current and high frequency, a significant perturbation of the voltage of the fundamental mode was observed, the effect of which was more pronounced in the lower energy RCSs due to the higher bunch charge. 
By choosing adjusted initial parameters at the start of the acceleration, injection transients of the voltage can be suppressed. 
To quantify the effect of the changes in beam phase, $\phi_b$ and cavity voltage, $V_c$ on the acceleration, a detailed beam dynamics simulation will be performed using the BLonD \cite{BLonD} simulation suite. 
\section{Acknowledgments}{
This work has been sponsored by the Wolfgang Gentner Programme of the German Federal Ministry of Education and Research (grant no.~13E18CHA), endorsed by the International Muon Collider Collaboration (IMCC), and funded by the European Union (EU). 
However, the views and opinions expressed are those of the authors and do not necessarily reflect those of the EU or the European Research 
Executive Agency (REA). Neither the EU nor the REA can be held responsible for them. Appears in the proceedings of the 14th International Computational Accelerator Physics Conference (ICAP’24), 2-5 October 2024, Germany. 
}
%
%
\ifboolexpr{bool{jacowbiblatex}}%
	{\printbibliography}%
	{	
	
} 

\end{document}